\documentclass[conference]{IEEEtran}
\IEEEoverridecommandlockouts
\usepackage{cite}
\usepackage{amsmath}
\usepackage{amsfonts}
\usepackage{amssymb}

\usepackage{threeparttable}
\usepackage{multirow}
\usepackage{algorithm}

\usepackage{graphicx}
\usepackage{textcomp}
\usepackage{xcolor}
\usepackage{subcaption}
\usepackage{algorithm}
\usepackage{algpseudocode}
\usepackage{bm}
\usepackage{esvect}

\DeclareMathOperator*{\argmin}{arg\,min}

\newcommand{\ele}{\vv{\bm{e}}}

\newcommand{\GF}{\mathsf{Goldfish}}
\newcommand{\VAR}{\mathrm{Var}}

\newcommand{\PLUS}{\dagger} 
\def\BibTeX{{\rm B\kern-.05em{\sc i\kern-.025em b}\kern-.08em
    T\kern-.1667em\lower.7ex\hbox{E}\kern-.125emX}}
\begin{document}

\title{Goldfish: Peer selection using Matrix completion in unstructured P2P network}


\author{\IEEEauthorblockN{1\textsuperscript{st} Bowen Xue}
\IEEEauthorblockA{\textit{Electrical and Computer Engineering} \\
\textit{University of Washington}\\
Seattle, US \\
bx3@uw.edu}
\and
\IEEEauthorblockN{2\textsuperscript{nd} Yifan Mao}
\IEEEauthorblockA{\textit{Computer Science and Engineering} \\
\textit{Ohio State University}\\
Ohio, US \\
mao.360@buckeyemail.osu.edu}
\and
\IEEEauthorblockN{3\textsuperscript{rd} Shaileshh Bojja Venkatakrishnan}
\IEEEauthorblockA{\textit{Computer Science and Engineering} \\
\textit{Ohio State University}\\
Ohio, US \\
shaileshh.bv@gmail.com}
\and
\IEEEauthorblockN{4\textsuperscript{th} Sreeram Kannan}
\IEEEauthorblockA{\textit{Electrical and Computer Engineering} \\
\textit{University of Washington}\\
Seattle, US \\
ksreeram@uw.edu}
}



\maketitle

\begin{abstract}

Peer-to-peer (P2P) networks underlie a variety of decentralized paradigms including blockchains, distributed file storage and decentralized domain name systems. A central primitive in P2P networks is the peer selection algorithm, which decides how a node should select a fixed number of neighbors to connect with.
In this paper, we consider the design of a peer-selection algorithm for unstructured P2P networks with the goal of minimizing the broadcast  latency.
We propose $\GF$, a novel solution that dynamically decides the neighbor set by exploiting the past experiences as well as exploring new neighbors. The key technical contributions come from bringing ideas of matrix completion for estimating message delivery times for every possible message for every peer ever connected, and a streaming algorithm to efficiently perform the estimation while achieving good performance.
The matrix completion interpolates the delivery times to all virtual connections in order to select the best combination of neighbors. $\GF$ employs 
a streaming algorithm that only uses a short recent memory to finish matrix interpolation. 
When the number of publishing source is  equal to a node's maximal number of connections, 
$\GF$ found the global optimal solution with $92.7\%$ probability by exploring every node only once. In more complex situations where nodes are publishing based on exponential distribution and adjusting connection in real time, we compare $\GF$ with a baseline peer selection system, Perigee\cite{mao2020perigee}, and show $\GF$ saves approximately $14.5\%$ less time under real world geolocation and propagation latency.

\end{abstract}


\section{introduction}
\label{system:introduction}
P2P overlay network is a fundamental layer required by any blockchain consensus protocol for block delivery and reaching consensus. In a proof of work blockchain like Bitcoin, a network with high delivery latency produces a higher chance of forking in the consensus layer, effectively reducing the security parameter for any adversary to overtake the honest chain \cite{dembo2020everything}. 
In a proof of stake blockchain like Ethereum, a high latency network can delay the block propogation, and possibly reduce the participating reward because of missing protocol assigned tasks. This loss can discourage home validators(nodes) from running their own hardware, and increase the barrier for decentralization. 
Decentralization is a key defense mechanism against censorship attack, which is required for the liveness property of any consensus protocol. Recently there was an outcry within Ethereum community on large staking pools engaging in censoring of certain transactions \cite{flashbot, thedefiant}. Such liveness concern can only be addressed by having a large set of decentralized nodes participating in the consensus protocol, which in turn would require underlying P2P protocol to be scalable. In the end, a node's safety also depends on security of the P2P layer. Inside the P2P network, if a node locally connects to the adversary only, its view on the state of the chain can be manipulated by feeding the wrong  blocks. Such an attack is called eclipse attack \cite{vyzovitis2020gossipsub}. Hence a P2P layer should secure nodes against connection manipulation. To summarize, a P2P protocol needs to address at least three aspects: latency, scalability and robustness to connection manipulation. We ask the question how we can design a broadcast P2P protocol that addresses those issues?
 


Suppose there are $N$ nodes in a P2P network, a fully connected graph is a design that has the most security. Its message delivery time is constant $O(1)$ and each node only receives the broadcast message only once. But such scheme
is not scalable since the total number of connections in the entire network is $O(N^2)$. One can design a structured P2P network like Kademlia\cite{maymounkov2002kademlia}, where each node only connects to $O(log(N))$ number of peers based on  consistent hashing. The network is scalable since each node only connects to a few nodes even within a large graph; it is also efficient since the network structure helps  disseminate messages such that there is no redundant message received by each node \cite{el2003efficient}. 
Besides Kademlia, another design that reduces latency is to use synthetic coordinates\cite{dabek2004vivaldi} for creating a structured network.
However, the structured design is not secure, since an adversary can always observe the network and pinpoint connections to compromise. Proposals have been made to modify structured P2P design to accommodate the security \cite{augustine2022fully}, but as far as authors' understanding such designs have not been adopted by  scalable blockchains.  

An unstructured p2p network is a balanced design between the above approaches, where connections are created by randomly choosing a constant number of nodes in the network. 
Such design is scalable because the node only connects to a constant number of peers. The security is also preserved since it is difficult for an adversary to predict and manipulate the node's local network given its connections are purely random out of all possible connections. The unstructured network has drawbacks of being less efficient because it relies on flooding its local network in order to broadcast the message; it is also more difficult to add latency optimization since there is no coordination among random connections from each node. However, in production, blockchains like Ethereum and Bitcoin both use the unstructured p2p design as its overlay network,  for its good scalability and safety properties. The particular unstructured network design used by Ethereum 2.0 is descrribed in\cite{vyzovitis2020gossipsub}. However, we still need to address the latency problem as it is a potential threat to the consensus protocol, and moreover, latency is an important factor in user experience for any protocol that wants to achieve fast time block finality.

Latency optimization in unstructured P2P networks is challenging, because an algorithm needs to be adaptable to variability introduced by network churn, and to the randomness from the publishing sources. Work has been done to treat the peer selection problem as a bandit problem \cite{mao2020perigee}  in which a node makes exploration and exploitation to minimize the latency. 
However, such scheme relies on a relatively simple heuristic that only considers the peer performance at the moment, and discard all the information explored in the past.
We noticed that there are still open design space inside peer selection problem, which improves latency while maintaining scalablility and security. Specifically, we can decouple the network modelling and peer selection into two parts, and we can optimize each of them with  higher design freedom. 

We propose $\GF$ that treats the peer selection as a learning problem whose data points are continuously collected from the network to create a model, which predicts network latency to any nodes. The model is able to adjust itself across time as the network experiences churn or randomness from the publishing sources.  
Specifically, a $\GF$ instance optimizes itself through four procedures: \textbf{data observation}, \textbf{matrix constructor}, \textbf{matrix completer} and \textbf{peer selector}. 
The \textbf{data observation} procedure takes care of collecting information from the network.
The input data 
requires only the IP address from the TCP header, without extra information from the P2P overlay protocol. 
Moreover, in the context of blockchain, $\GF$ can optimize connections to miners even if the block does not include timestamp of creation, which can provide additional privacy defense.

The core algorithm considers every input data point as a (peer, message, delivery time) tuple observed by a $\GF$ instance. Those data points are used by the \textbf{matrix constructor} to fill in a matrix whose dimensions are message and peer. The missing entries represent the unknown delivery time for some messages with some unconnected peers. A naive implementation would be to create a large matrix for every possible message for every peer ever connected. But completing such matrix would be computationally intensive and slow, preventing the system to adapt with the network changes in real-time. In addition, using data from old history could also leads to inaccurate results, because data points collected long time ago are not reflective of the current network state. 
We solve both the issues by using a streaming algorithm to digest information in real time and deprioritize old information. This design removes the constraint to interpolate the entire matrix, and is still able to attain optimal solution by incrementally finding the best peer within a local region. 

The key step for interpolating the missing cells is to solve a non-convex optimization problem derived from the matrix, the procedure is carried out by a \textbf{matrix completer} procedure
which uses an optimizer to find the best value such that the interpolated missing values 
are consistent to known value in the matrix. We use $K$ nearest neighbor (see Section~\ref{sec:goldfish}) to pick up the consistent peers.
After interpolating the missing values, the \textbf{peer selector} procedure uses an altruistic exploitation strategy by choosing best peers from the existing set of peers that minimizes the broadcast latency. To dynamically adapt to the network as well as to explore new peers, a $\GF$ establishes new connections to random peers every time the exploitation strategy is run. Each node can only have a constant number of outgoing connections, the exploration process requires to drop some existing peers. But since exploration in nature is random, sometimes a good performing peer is replaced by a random peer who turned out to be less performant.
$\GF$ inherently has the memory about which peers had good performance in the recent history, and therefore re-catches the good performing peers if they were dropped during exploration. 

To validate the design of $\GF$, we create a simulation tool that measures latency among any pair of nodes in a network. The network can be generated with arbitrary size with ability to specify constraints on degree of any node. Every node assumes two roles. A node must be either a publishing or forwarding node for the first role; and a node must either be a static node that fixes its peers, or a adaptive node that actively selects its peers  for the second role.
The broadcast latency measures how well a node is choosing its peers, and is computed efficiently using the shortest path algorithm. 
We evaluate $\GF$ under both analytical and complex dynamic settings. In the analytical setting, only 1 $\GF$ node is adaptating and the number of publishing source is equal to the degree of the node.  We can quantify the performance by clearly defining success and failure: if $\GF$ can attain the global optimal solution 
which is direct connections to all publishing sources. We show that in a network size of 100 nodes with 3 publishers, a $\GF$ node succeeds with $92.7\%$ probability from 200 randomly generated network by only exploring every peer only once. In more complex dynamic settings, many independent nodes are publishing and adapting simultaneously. Because the global optimal solution for all node simultaneously is computationally intractable, we compare $\GF$ with a baseline algorithm Perigee\cite{mao2020perigee} to solve the same problem. We show that $\GF$ has strictly superior performance in all cases, and reduces broadcast latency by  $14.5\%$ than the baseline algorithm. 

Section~\ref{sec:related} summarizes the related works. Section~\ref{sec:system_model} describes data measurement methods and network assumption. Section~\ref{sec:goldfish} dives into details of $\GF$. Section~\ref{sec: sim_and_exp} presents the simulation tool and experiment results. The code is available at \cite{goldfish}.

\section{Related work}
\label{sec:related}
\subsection{P2P network and Perigee}
A P2P network is an overlay logical network that has a wide range of applications including distributed file storage\cite{ripeanu2001peer}, distributed lookup table\cite{maymounkov2002kademlia}. Because connections on an overlay network are logical, the physical property like peer-to-peer distance is abstracted away, therefore the network topology has great impact on a node's shortest path to other nodes in the network. 
Perigee\cite{mao2020perigee} is a system specifically built for blockchain, it treats peer selection as a tradeoff of exploration vs. exploitation. It uses local time measurement similar to $\GF$  as the only source of information for making peer selections. However, unlike $\GF$, Perigee only uses current set of connections to decide which peers to keep. Its core (called subset) algorithm  enumerates all possible combination of exploitation subset from the current peers, then associates each combination with a score generated using  time observation in its current connections; it chooses the exploitation peers which achieves the best score. Perigee was shown to have similar, if not superior, performance than Kademlia in many network scenarios.

\subsection{KNN and tensor graph}
The Matrix completion problem in $\GF$ can be categorized as a unsupervised learning problem, whose input data does not contain any label and is often represented in vector form. $\GF$ employs K-nearest neighbor (KNN) to solve this learning problem, which is a method that looks for using $K$ closest vector surrounding the interested vector for interpolation. Since the optimization problem in $\GF$ is non-convex, we use  Pytorch \cite{pytorch} to empirically evaluate variables for finding their local optimal solution. Pytorch is an optimized tensor library for deep neural network, which has shown been effective to train machine learning model with appropriate loss functions. 


\section{System Model}
\label{sec:system_model}
A node inside a P2P network can be an adaptation node, a publishing node or both. An adaptation node continuously optimizes its connections, as opposed to a static node which always keeps the same set of connections. A publisher node generates original messages depending on its publishing probability. In blockchain, publishers are the miners, and adapting nodes are full nodes that optimize their outgoing connections.

Every node has constraints of CPU and bandwidth, which translate to how many concurrent peers a node could maintain connections. In this work, we assume a model that every node can at most maintains a constant number of outgoing and incoming connections. A node has 2 actions on how to react to an incoming connection, by either rejecting it if the node has saturated all its connections or by accepting it. A node can freely choose any  peer combinations for initiating its outgoing connections, though some outgoing connections might be rejected by the other node if the other node is saturated. In this work, we assume each node knows and has means to connect to all other nodes in the system, and therefore a $\GF$ needs to make decisions from an exponential number of combinations of connections.

Every P2P node is required to relay messages. Any connection is bidirectional, and every node is required to relay messages for all incoming and outgoing connections, except to the peer from whom the node firstly received the message. All nodes are expected to receive multiple copies of the identical messages from all its peers. We do not assume a message to have timestamp or any other identification related to who generates the message. All nodes in the network need not synchronize their time. Instead, each node uses relative time measurement for each message to record its peer's performance. Suppose peer $A$ sends a message, and the message is the first one that arrives at the local node at local time $t$, then peer $A$ has a measurement of $t-t=0$; peer $B$ sends the same message at $t'$, and has a measurement of $t'-t$ in local time. Such measurement is easy to implement and robust to time tempering since it only relies on local observation.


\section{Goldfish}
\label{sec:goldfish}

\begin{figure}[hbt]
    \centering
    \includegraphics[width=0.42\textwidth]{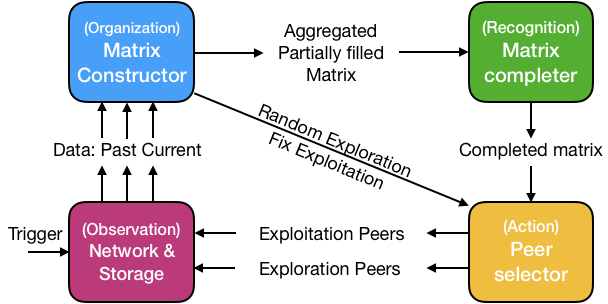}
    \caption{Goldfish components flowchart}
    \label{fig:pipeline}
\end{figure}

$\GF$ is a distributed algorithm that each node runs locally to optimize its  
shortest topological distances to other nodes in the P2P network. $\GF$ mimics an intelligent entity optimizing for its own improvements: it collects surrounding information, organizes it and recognizes pattern in order to make next move to fit  in a ever-changing environment. They correspond to four components, as shown in Fig~\ref{fig:pipeline}. They are Network-Storage, Matrix Constructor, K-NN Matrix completer and Peer Selector. 

We refer to the data forwarded by a node as a message, and refer to the content of a message as a block. A node can receive multiple messages of the same block. The Network-Storage component observes and keeps a memory about any delivered data in the local storage and formats it as a (peer, block, delivery time) tuple. The delivery time is measured with relative timestamp, see Section \ref{sec:system_model}.

At initial, $\GF$ is triggered to collect data tuple, and uses a Matrix Constructor to organize observations from the local storage for creating data batches. A batch contains messages observed in a contiguous time span without connection changes, we use epoch $E_1,..E_\ell$ to denote each of the $\ell$ batches 
corresponding to each contiguous time span.

For messages collected in a single epoch, we can create a 2D matrix, where each block has its own row vector, and each peer has its own column vector. The intersection of row and column stores the recorded latency. But to concatenate blocks from multiple epochs into a single matrix, every block vector has to insert empty cells for peers only present in other epochs. The \textbf{Matrix Constructor} creates a synthesized matrix by concatenating epochs along the block axis, and its peer axis is union of connections across epochs. A synthesized matrix from 2 epochs is shown in Table(\ref{table:constructor}) where $\star$ are missing observations, which is further discussed in the following sections.

$\GF$ uses Matrix completer to estimate the missing cells. First, the completer classifies all cells in the matrix into 4 categories: \textbf{observed}, \textbf{symbolically observed}, \textbf{estimable} and \textbf{inestimable}. For all estimable missing cells, they are formulated as optimization variables in a non-convex optimization formulation. While designing the loss function, we have a key insight that block vectors exhibiting similar peer delivery pattern are more likely from the source-nodes close to each other. We can compare those block(row) vectors and use K-NN to interpolate the missing cell using nearest neighbor. To solve for solution, we create a pytorch tensor graph (see Section~\ref{sec:related}) to empirically optimize for the missing values. 

After estimating the missing cells,
$\GF$ can use any peer exploration algorithm. One particular selection algorithm that helps to evaluate $\GF$ performance is the depleting pool, where exploration peers are drawn from a depletion pool which includes all nodes in the network at the beginning; the peers are taken out of the pool without replacement, and resets itself until the pool is emptied. 
Exploitation peers are chosen based on an altruistic heuristics. We first identify a set of contributing peers who always deliver blocks fast to us, and we credit them scores by counting the number of benefiting peers who receives block fast from us. The altruistic principle is based on an intuition: if a $\GF$ can help so many other nodes, then itself must already be well connected, so a $\GF$ should improve itself by the measure how helpful it becomes to the others. A $\GF$ can occasionally skip recognition step if it wants to speedup exploration as represented by the diagonal in Fig.~\ref{fig:pipeline}, and this is implemented inside Scheduler submodule.

Since the system is remarked by its short memory only to appreciate recent data (because historic data is not reflective to current network condition), the completer only needs to process a few epochs at a time. It greatly reduces the space and computation complexity. 
For the rest of the section, we characterize every component of $\GF$ with greate details.
 


\begin{table*}[!hbt]
\renewcommand{\arraystretch}{1.35}
\centering
\caption{Transformation from (a) Synthesized matrix to (b) Output of Matrix completer }
\begin{subtable}[h]{.385\linewidth}
\resizebox{\textwidth}{!}{
\begin{tabular}{|l|l| c c c c|}
\hline
 \multirow{2}{*}{}  &  Block    & \multicolumn{4}{c|}{Peers} \\
   &   & $n_1$ & $n_2$ & $n_3$ & $n_4$\\ 
\hline
\multirow{3}[0]{*}{ $E_1$ } &
  $m_1^{n_5}$ & $0$       & $t_{1,2}$ & $\star$ & $t_{1,4}$\\
& $m_2^{n_2}$ & $\PLUS$   & $0$       & $\star$ & $\PLUS$\\
& $m_3^{n_5}$ & $0$       & $t_{3,2}$ & $\star$ & $t_{3,4}$\\
& $m_4^{n_6}$ & $t_{4,1}$ & $t_{4,2}$ & $\star$ & $0$\\
\hline
\multirow{3}[0]{*}{ $E_2$ } & 
  $m_5^{n_2}$ & $\PLUS$   & $0$   & $\PLUS$ & $\star$  \\
& $m_6^{n_6}$ & $0$       & $t_{6,2}$ & $t_{6,3}$ & $\star$ \\
& $m_7^{n_6}$ & $0$       & $t_{7,2}$   & $t_{7,3}$ & $\star$\\
& $m_8^{n_7}$ & $\dagger$ & $0$ & $t_{8,3}$ & $\star$\\
\hline
\end{tabular}
}
\caption{$\star$ is missing cell. $\dagger$ is symbolically known cell when a peer gets the first block from local node }
\label{table:constructor}
\end{subtable}
$\Longrightarrow$
\begin{subtable}[h]{0.548\linewidth} 
\resizebox{\textwidth}{!}{
\begin{tabular}{|l|l| c c c c|}
\hline
 \multirow{2}{*}{}  &  Block   & \multicolumn{4}{c|}{Peers} \\
   &   & $n_1$ & $n_2$ & $n_3$ & $n_4$\\ 
\hline
\multirow{3}[0]{*}{ $E_1$ } &
$m_1^{n_5}$ & $c_1$       & $c_1+t_{1,2}$ & $a_1$ & $t_{1,4}+c_1$\\
& $m_2^{n_6}$ & $\PLUS$   & $c_2$       & $\times$ & $\PLUS$\\
& $m_3^{n_6}$ & $c_3$     & $t_{3,2}+c_3$ & $a_2$ & $t_{3,4}+c_3$\\
& $m_4^{n_6}$ & $t_{4,1}+c_4$ & $t_{4,2}+c_4$ & $a_3$ & $c_4$\\
\hline
\multirow{3}[0]{*}{ $E_2$ } & 
  $m_5^{n_2}$ & $\PLUS$ & $c_5$   & $\PLUS$ & $\times$  \\
& $m_6^{n_6}$ & $c_6$   & $t_{6,2}+c_6$ & $t_{6,3}+c_6$ & $a_4$  \\
& $m_7^{n_6}$ & $c_7$   & $t_{7,2}+c_7$ & $t_{7,3}+c_7$ & $a_5$ \\
& $m_8^{n_7}$ & $\dagger$ & $c_8$   & $t_{8,3}+c_8$ & $\times$\\
\hline
\end{tabular}}
\caption{Completed matrix. $\times$ is infeasible cell. $\dagger$ is symbolically known cell. $\ddagger$ is ambiguous cell. $a_1-a_5$ are estimated missing values. $c_1-c_8$ are offset.}
\label{table:completer}
\end{subtable}

\end{table*}

\subsection{Matrix Constructor}
Matrix constructor transforms the tuple data 
from multiple epochs to a single partially observed matrix. A concatenated matrix with 2 epochs in shown in Table (\ref{table:constructor}). In both epochs, the number of active connections is 3. In epoch $E_1$, the node is connecting to peers $n_1, n_2, n_4$; whereas in epoch $E_2$, the node is connecting to peer $n_1, n_2, n_3$. There are 4 blocks in each epoch being delivered to the node, with notation $m_1, ..., m_8$. Each superscript on the block denotes the original publisher for the block, which is used only to elaborate the algorithm, and 
$\GF$ does not need this information for execution.
As shown in Table(\ref{table:constructor}), after combining 2 epochs, each epoch contains a missing sub-column because those peers are not connected when the corresponding blocks were perceived. In the next step, $\GF$ categorizes cells in order to formulate the optimization problem.


\subsubsection{Preliminary Cell classification}
A partially observed matrix, $T$, of size $(p,q)$ contains 3 types of cells: observed, missing, and symbolically known cells. Every cell in the matrix is associated with 2 attributes: the delivering peer and the delivered block. A cell $(i,j)$, where $0<i\le p, 0<j\le q$, is an observed cell if there is a peer $n_i$ delivered a block $m_j$ to the node, and $t_{i,j}$ is the associated measurement. The measurement for an observed cell $(i,j)$ is computed as 
time difference between the peer $n_i$'s delivery time and the earliest delivery time for block $m_j$ among all $q$ peers. 
If a peer $n_i$ does not send the block $m_j$ to the node, the cell $(i,j)$ is categorized either as a missing cell($\star$), or as a symbolically known cell($\dagger$). 
A missing cell is identified if the cell is synthetically created as the result of merging multiple matrices along the block axis; a symbolically known cell is identified when the cell's associating peer does not forward the block to the $\GF$ node; 
in which case, we can infer the local node (who is creating the table) is the first node who forwards the block to that peer (since a node should not send back the block to where it comes from). 
For example in Table(\ref{table:constructor}), cell $(2,1)$ and $(2,4)$ are symbolically known, and we can infer the local node is the first one that is delivering the block $m_2$ to peer $n_2, n_4$. 
The table suggest a situation where $n_1, n_3, n_4$ are far away from the original publisher $n_2$ for $m_2$, and the local node is on the faster path to deliver the block to other peers. 
Although the symbolically known cells are missing measurement data, they still worth to consider because they convey the associated peer cannot be the fastest ones to deliver similar blocks to the local node.
These 3 cell types partition $p \times q$ cells, and each cell type induces its own binary masking matrix on $T$. 

\subsubsection{Scheduler}
A local node can combine arbitrary number of epochs into a matrix for cell interpolation. But as network continues to change, historical data might damage the completion accuracy because 
the network topology might have changed so much that those date are no longer useful to represent the current network. There are 2 design spaces to tune learning behavior: the size of the synthesized matrix determined by the number of epoch, and frequency to initiate the algorithm. For example, a node can set the matrix to contain 3 epochs, and choose to run matrix completion every 2 epoch. In which case, instead of running completion every time when peers change, the node can skip 1 learning epoch and use it to explore more peers. This speeds up node exploration, and saves resources on running computational tasks.


\subsection{K-NN Matrix Completer and Missing Cell Classification}
After data is formatted into a matrix, $\GF$ uses K nearest neighbor to interpolate the missing cells. We begin by treating every row of the synthesized matrix, $T$, as a vector in the space $\mathbb{R}^q$, ($T_{i,j} = m_i[j]$), the goal is to cluster similar rows and use the available data to fill the missing ones.


\subsubsection{Estimable Missing Cells}
The missing cells are further classified into 3 categories: estimable, 
ambiguous ($\ddagger$) and infeasible($\times$). A missing cell attributed by block $r$ peer $u$, $m_r[u]$, is estimable if there exists another block $m_i$ which has measured data from peer $u$, and  both blocks $m_i, m_r$ have measured data from least 2 common peers. The intuition is that, we can use the closeness of the observed value to derive the unobserved value.
But since we use K nearest neighbor to interpolate the missing dimensions, we need to find  K such blocks to complete the estimation. However, not all missing cells can find enough $K$ blocks 
to satisfy the requirements. Those cells are deemed as ambiguous ($\ddagger$). Infeasible cells are missing values that cannot be estimated, its definition is not central to peer selection, so we move it to the end.

\subsubsection{Distance Metrics for Selecting K Nearest Neighbor}
\label{example}
A metric function $g: \mathbb{R}^p \times \mathbb{R}^p \rightarrow \mathbb{R}^+$ 
takes two block vectors and produce a non-negative value.
A metrics is good if it produces a small value when   
two block sources are close and they have similar delivery viewed from a perspective of the local node.
Let's first focus on a simple case where all publishers have perfect synchronous clocks and block creation time is printed on all messages. Then if a node receives multiple messages about the same block, it can deduce about its peers' topological positions in the network based on 
the messages arrival time. 
But in the real world, we cannot get accurate time measurement because time synchronization precision among computer can be low, and bad data can significantly damage the model. In Section~\ref{sec:system_model}, we discussed to use relative time to solve the problem, and it has an effect to offset the fastest route to 0. It appears we miss some information, because all values subtracts a common constant, but subtracting a offset does not change the differential value among vector's dimensions. Therefore, we can still use the differential characteristics of the block vector to infer if 2 block vectors are coming from close publishers.
To construct the distance metrics, 
we subtract 2 vectors and take the unbiased variance on the peer dimensions where the local node has observations to both blocks, i.e. $g(m_r, m_i) = \VAR(D_i)$ where $D_i = \{m_r[j] - m_i[j]: 0<j\le q, B_{r,j}=1, B_{i,j}=1\}$ ($B$ is a binary matrix indicating if a cell has observation). 
At least 2 values are required to compute unbiased variance. 

The differential variance metrics has a good property that groups blocks whose miners are close to each other. The intuition behind is that publishers with close topological locations generate blocks that exhibit similar differential variance. We illustrate this property by an example. From the Table(\ref{table:constructor}), since $m_4[4] = 0$ we know $n_4$ is close to $n_6$  compared to other nodes, such that $n_4$ is the first one to deliver $m_4$ (mined by $n_6$) to the local node, but $n_4$ is not fast enough to make the local node be the first forwarding node to $n_1, n_2$, i.e. $n_1, n_2$ have their separate but slightly worse path than $n_4$ relative to the miner $n_6$. In epoch $E_2$, peer $n_4$ is dropped by the local node  to connect with $n_3$. 
Suppose this is the only connection change in the entire network, then relative time difference for $n_1, n_2$ to deliver the block is identical as before, i.e. $t_{4,2} - t_{4,1} = t_{6,2} = t_{7,2}$. The differential variance metrics $g(m_4, m_6)$ is computed as $\VAR(\{t_{4,1}, t_{4,2}-4_{6,2}\}) = \VAR(\{t_{4,1}, t_{4,1}\}) = 0$,  and similarly $g(m_4, m_7) = 0$. Therefore $m_4, m_6, m_7$ are grouped together in the next section for optimizing the missing cells.

\subsubsection{Optimization Formulation.}
After defining the optimization space $\mathbb{R}^q$ and the metric to find $K$ nearest neighbors, we start to formulate the unsupervised optimization problem.
Suppose there are in total $s$ estimable missing values, whose set is denoted as $S$. In the partially observed matrix $T$ of dimension $p\times q$, we associate each missing value with an optimization variable, and denote them as $A =a_1, ... , a_s$.
Each missing value in the matrix has a row and column indices, which can be encoded by 2 functions: 
$
    \textbf{row}: [s] \rightarrow [p] \quad \textbf{col}: [s] \rightarrow [q] 
$. Notation $[n]$ denotes a integer set $\{1,..,n\}$.
For each missing cell $a_v \in S$, its $k$ nearest vectors can be pre-processed using the metrics defined earlier, and the processed result can be summarized by 2 functions. Function $\textbf{N}: [s] \rightarrow [p]^k$ encodes for each missing cell $a_v$, which $k$ block vectors out of $p$ block vectors are closest to the block $m_{row(a_v)}$ where $a_v$ is a part of. Furthermore, the loss calculated by metrics $Var(D)$ between $m_{row(a_v)}$ and any of its $k$ nearest vector can be encoded as a function $\textbf{w}: [s, p] \rightarrow \mathbb{R}$. Since we only assign weight to $k$ such vector, other vectors has a weight of $0$. We further normalize the $k$ weight so that their sum equals to 1 using softmax \cite{softmax}. We also define a helper function $\textbf{L}: [p, p] \rightarrow \{0,1\}^q$ which takes 2 integers $i,j$ and returns a $q$ dimensional binary vector indicating peers who deliver block $m_i, m_j$ to the local node.

To illustrate how to construct the loss function, we begin with a simpler task by optimizing for a single missing cell $a_v$. We can get its $k$ closest vectors and the corresponding weight using the function defined earlier. $N(v)$ gives $k$ peers and $w(row(v),o)$ gives their weights. Since all latency are measured with relative time, we have a problem that any 2 row vectors are not comparable to each other due to a lack of common reference time. To illustrate it, 
we performed a step-by-step cells completion 
for both blocks $m_4$ and $m_6$ in the Table(\ref{table:constructor}). 
Suppose $m_4, m_6$ are  grouped together and  they are generated from one publisher,
we can calculate the vector difference based on measured time and use that to complete the missing ones. In this case, the difference are $t_{4,1} - 0 = t_{4,2} - t_{6,2}$.
As the result, the missing cell $m_4[3]$ should equal to $t_{4,1} + t_{7,3}$ and the missing cell $m_7[4]$ should equal to $- t_{4,1}$. 
As we can see, although we can compute those missing values, we are not able to compare latency across rows. Specifically block  $m_7$ is slower than $m_4$ by $t_{4,1}$, but the negative value in $m_7[4]$ is the lowest number.
Consequently, it is much more helpful if there is a common reference such that we can easily compare latency among multiple row vectors based on absolute number. For that purpose, we introduce extra optimization variable, $C = c_1, ..., c_p$ for each row. We show an example of a compensated matrix with all optimization variables in Table(\ref{table:completer}). With all variable defined, the  optimization formulation is shown below
\begin{align}
    \argmin_{A, C} & \sum_{v = 1}^{s}
    \sum_{o \in N(v)} w(v,o) \| I( \Delta_{v,o} , \Theta_{v,o} ) \|_2 
                        +\|A\|_2 + \|C\|_2  \label{eq1} \\ 
	\textrm{s.t.} \quad 
	                & \Delta_{v,o} = L(row(a_v), o) + \ele_{col(a_v)} \label{eq2}\\
				    & \Theta_{v,o} = T_{o} + c_{o}\mathbf{1}^q - T_{row(a_v)} - c_{row(a_v)}\mathbf{1}^q \nonumber\\
                    & \quad \quad \quad  - (a_{v}  - c_{row(a_v)}) \ele_{col(v)} \label{eq3}
\end{align}
We use this formulation to continue to optimize for a single cell $a_v$. After we identify $k$ nearest vectors and their weight, we optimize for the $\ell_2$ norm on the difference between the 2 vectors. The vectors subtraction is shown in equation \ref{eq3}, where $T$ is a matrix containing all observed latency data, and all unobserved data are filled with 0; notation $T_{o}$ is the $o$-th row of the matrix $T$. As shown in the equation \ref{eq3}, $\Theta_{v,o}$ is the difference between the row vector $m_{row(v)}$ where $a_v$ comes from, and the neighboring vector $o$.
But since each of 2 vectors contains many other missing cells, and those cells do not contribute to the estimation of $a_v$. They are not considered in the metrics weight calculation. It is done by creating an indicator vector $\Delta_{v,o}$ in equation \ref{eq2} with the helper function $L(row(a_v), o)$ and an elementary vector $\ele_{col(v)}$ to specify which dimensions should be contributing to the loss.
We enable it with a mask selector function $I(a, b) = \{b_j: 0<j\le q, a_j=1\}$ function to combine equation \ref{eq2} and \ref{eq3} in equation \ref{eq1}. To optimize for all optimization variable, we optimize for sum of loss and add two regularization terms to keep optimization solution bounded.


\begin{figure*}[hbt]
    \centering
    \includegraphics[width=0.75\textwidth]{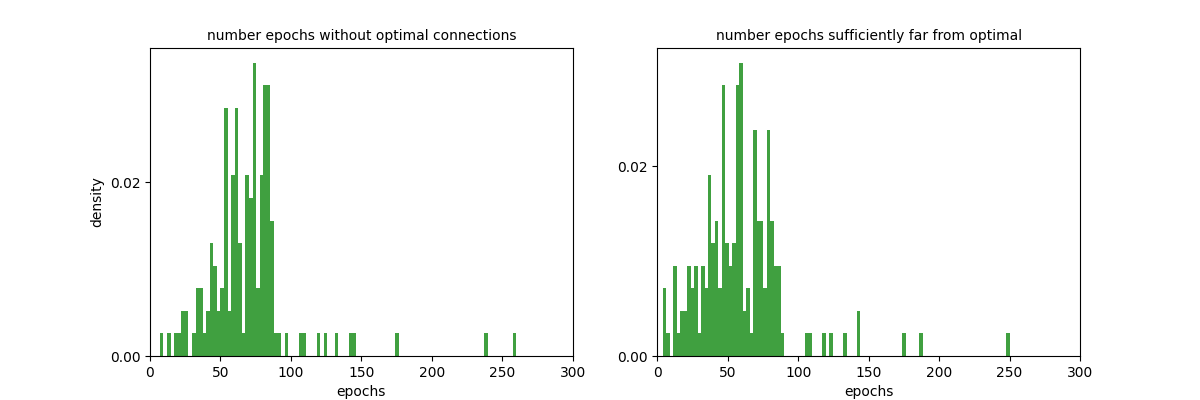}
    \caption{Histogram of number of non-optimal epochs}
    \label{fig:optimal}
\end{figure*}

\subsubsection{Optimization Solver.}
The stated optimization problem is an unconstraint non-convex optimization problem, which is difficult to solve by analytically deriving gradients for each variables. Instead, we use a popular pytorch autograd package proven to be successful in many machine learning tasks. See Section~\ref{sec:related} for pytorch tensor graph. Solving the formulated problem requires 3 components: node definitions, a loss function that defines graph edges, and an optimizer. 
We define $A, C$ as the edge nodes that requires gradients and each observation row vector $T_i$ as constant nodes, so that the back-propagation only optimizes variables $A,C$. The loss function defined in the last section gives straightforward instructions to construct DAG tensor graph: the binary vector $\Delta_{v,o}$ decides edges for the graph, and the computation logic is contained in $\Theta_{v,o}$; in the end, we regularize $A, C$ by adding them to the final cost. Since  none of variables $A, C$ is constrained, we can use optimizer provided from pytorch like adam, which has been highly optimized, and gives us a better performance in short time. It is important to balance the complexity of edges among tensor nodes. 

\subsubsection{Infeasible Missing Cells}
A missing cell can be infeasible ($\times$) if its value cannot be estimated, which can occur in 2 situations. First, when there is only 1 numeric observation in vector $m_r$. 
It happens when the local node is the fastest relaying node. 
In the Table(\ref{table:constructor}), cell $m_2[3], m_5[4]$ are infeasible. Second, a missing cell, $m_r[u]$, is infeasible if there is no  vector $m_i$ satisfying 2 properties: the local node has numeric observation for block $i$ from peer $u$; the local node has numeric observations for block $i$ and $r$ from at least 2 peers (number of 2 is discussed in Distance metrics).

\section{Experiment and Evaluation}
\label{sec: sim_and_exp}

We aim to investigate the following questions through evaluations: in a random network (1) Can a single $\GF$ instance use a short memory to find and retain the global optimal connections quickly when the unique global optimal solution is direct connections to all publishers? (2) What is the performance of multiple simultaneous $\GF$ running under varying number of publishers when global optimal for all adapting nodes is computationally intractable? (3) What is the performance of multiple $\GF$ in networks with real world propagation latency under varying number of publishers? 
We address (1-3) with a simulation tool.

\subsection{Simulation Framework}
For answering (1) and (2), we create multiple artificial networks inside a 2D plane, where locations of nodes are uniform randomly generated along both axes. For answering (3), we use measured latency\cite{wonderproxy} among randomly selected cities in the real world to create a 3D network graph. Every node is constraint by $8$ incoming connections and $4$ outgoing connections; any pair of nodes need to respect both constraint before setting up a connection. Every node needs to relay messages as specified in Section~\ref{sec:system_model}. Network operates in round fashion. Only one publisher generates one message in each round, and each node has a fixed publishing probability, together sum to 1. Every message is broadcast until every node has a copy. The delivery latency between 2 directly connected peers has two components: fixed 20ms node (including processing and transmission) delay and propagation delay. The latency of delivering a message between any 2 nodes is the shortest path on the weighted graph, whose edge weight is propagation latency. We use Dijkstra algorithm to find the shortest path using only the exploitation edges. Each experiment is initialized with a random network topology (respecting the edge constraints) using a random seed. Every $\GF$ node uses $3$ outgoing connections for exploitation and $1$ for exploration. Every epoch contains $40$ messages(rounds), and the matrix constructor combines $3$ epochs, where the middle epoch is always the pure exploration epoch, see Section~\ref{sec:goldfish} Scheduler. Matrix completer sets $K=2$ for nearest neighbor problem, and runs at most 2000 optimization steps. The typical amount of time to run a matrix of size $120 \times 7$ requires around $20$ seconds on a 6 cores 2.6G Hz 16 memory laptop.

\begin{figure*}[!hbt]
    \centering
    \includegraphics[width=0.73\textwidth]{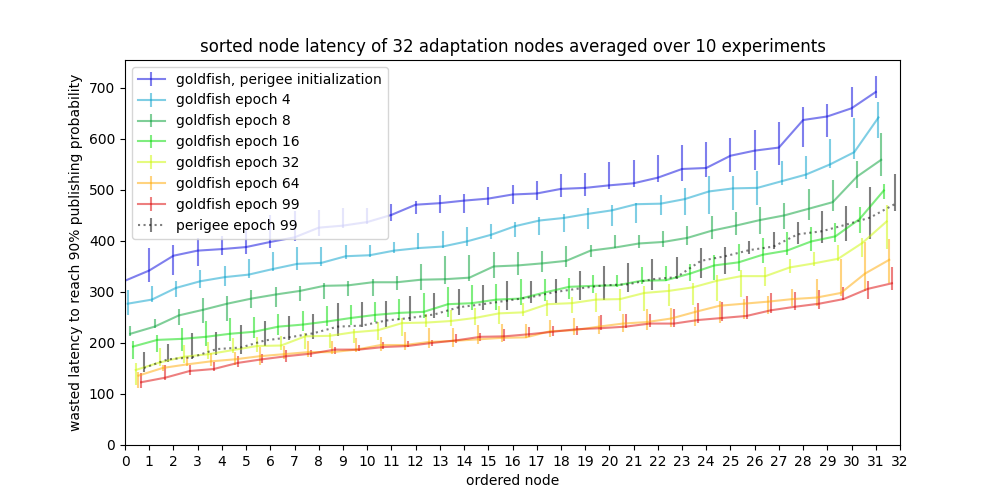}
    \caption{Performance comparison between Goldfish and Perigee in a random graph of 100 nodes with exponential publishing probability. Solid line connects  median values, upper bar shows the 75th wasted latency, and bottom bar for 25th of 10 experiments. The dashed line shows captures the network that uses Perigee.}
    \label{fig:goldfish_perf}
\end{figure*}

\begin{table*}[hbt]
\centering
\caption{$\GF$ Perigee Latency on 10 experiments under both random and real model. measured in millisecond}
\scalebox{0.94}{
\begin{tabular}{|c|c|c| c | c | c | c| c| c| c| c| c|c|}
\hline
topo & $\#$ & pub & $\#$ & \multicolumn{4}{c|}{$\GF$}  & \multicolumn{4}{c|}{Perigee} & ratio  \\ \cline{4-12} type & pub & prob & adapt & 25th & 50th & 75th & $mean$ & 25th & 50th & 75th & $mean$ & mean \\ 
\hline
\multirow{10}[0]{*}{ random } &  \multirow{4}[0]{*}{ $100$ } & \multirow{4}[0]{*}{ exp }
   & 10 & 249 & 288 & 342 & 299 & 262 & 319 & 370 & 323 & 0.926\\
 \cline{4-13}
 & & & 32 & 181 & 218 & 259 & 222 & 237 & 296 & 356 & 306 & 0.725\\
 \cline{4-13}
 & & & 64 & 155 & 185 & 217 & 191 & 188 & 227 & 270 & 235 & 0.813\\
 \cline{4-13}
 & & & 100& 134 & 162 & 193 & 169  & 151 & 173 & 201 & 178 & 0.949\\
\cline{2-13}
& \multirow{3}[0]{*}{ $32$ } & \multirow{3}[0]{*}{ unif } 
   & 10 & 261 & 294 & 354 & 308 & 288 & 315 & 381 & 332 & 0.928\\
 \cline{4-13}
 & & & 32 & 197 & 228 & 264 & 234 & 255 & 293 & 330 & 298 & 0.785\\
 \cline{4-13}
 & & & 64 & 164 & 188 & 218 & 194 & 189 & 218 & 260 & 229 & 0.847\\
 \cline{4-13}
\cline{2-13}
 & \multirow{3}[0]{*}{ $64$ } & \multirow{3}[0]{*}{ unif }
   & 10 & 287 & 320 & 358 & 328 & 316 & 348 & 390 & 358 & 0.916\\
 \cline{4-13}
 & & & 32 & 250 & 282 & 312 & 284 & 260 & 301 & 342 & 306 & 0.928\\
 \cline{4-13}
 & & & 64 & 195 & 218 & 247 & 226 & 195 & 221 & 257 & 229 & 0.987\\
\cline{1-13}
 \multirow{4}[0]{*}{ real } & \multirow{4}[0]{*}{ $100$ } & \multirow{4}[0]{*}{ exp }
   & 10 & 130 & 154 & 188 & 162 & 144 & 181 & 218 & 184 & 0.880\\
 \cline{4-13}
 & & & 32 & 114 & 131 & 158 & 140 & 128 & 152 & 193 & 165 & 0.848\\
 \cline{4-13}
 & & & 64 & 100 & 114 & 131 & 120 & 118 & 136 & 163 & 145 & 0.828\\
 \cline{4-13    }
 & & & 100& 86 & 99 & 114 & 104 & 98 & 111 & 130 & 120 & 0.867\\
\hline
\end{tabular}}
\label{table:table:rand_g_perf}
\end{table*}

\subsection{Global Optimal}
We claimed that $\GF$ can use short memory to find and retain the best connections with high probability only by exploring every peer once. To evaluate the idea, we setup experiments where the number of publishers is less than or equal to exploitation limit; in such situation, the global optimal solution is direct connections to all publishers. The experiments contain 200 random graphs on a $500 \times 500$ plane, and each experiment runs on a distinct graph configuration of random node locations and initial network topology for $300$ epochs. In each graph, 3 random nodes out of 100 nodes are assigned with publishing probability $\frac{1}{3}$, and 1 other node is randomly selected to run $\GF$ for changing outgoing peers (3 exploitation and 1 exploration) per epoch; all other nodes remain static. Note that a $\GF$ has to spend at most $96$ epochs to discover every peer once. 

Fig.~\ref{fig:optimal}(a) is a histogram generated from the $200$ random experiments whose x axis is the number of epochs that a $\GF$ node does not attain optimal solution. This condition captures both the convergence rate to find the optimal solution and tendency to diverge from the optimal solution. It shows that in most (around $92.7\%$) cases $\GF$ can find and retain the optimal solution by exploring every peer only once.

Fig.~\ref{fig:optimal}(b) is a similar histogram that depicts how many epochs the $\GF$ node is sufficiently far away from its unique optimal solution. Let $\lambda_i(e)$ be the latency difference at epoch $e$ between the optimal solution and its current Dijkstra distance; we compute and plot the sum of their difference for every epoch every experiment $\lambda(e) = \sum_{i=1,2,3} \lambda_i(e)$; the criteria for an epoch to be sufficiently far is determined as $\lambda(e) / \lambda(0)>0.05$, where $\lambda(0)$ is latency which a $\GF$ node gets from random connections. From the figure, the distribution is shifted downward, around 61.5\% cases $\GF$ finds sufficiently close solution within 48 epochs. 


\subsection{Multiple $\GF$ on Many Publishers Network}
When a $\GF$ is optimizing for more publishers greater than its exploitation limit, the node needs to compromise connections to balance latency to all publishers. In such complex system, finding the global optimal solution is intractable, hence we evaluate performance by comparing $\GF$ against a baseline algorithm Perigee\cite{mao2020perigee}, see Section~\ref{sec:related}.
Figure~\ref{fig:goldfish_perf} is a comparison for a 100 nodes random graph which contains 32 adapting nodes, 68 static nodes, and all nodes have publishing probabilities based on an exponential distribution (80\% probability is concentrated on 20 nodes). For evaluating broadcast latency of each node, we use only exploitation edges to compute the weighted shortest distance to all publishing sources. The broadcast latency is identified as the distance to reach $90\%$ of the publishing probability. To visualize how much extra time wasted on the topological relay as opposed to direct connection, we subtract each node's topological latency with their direct connection latency, and plot the sorted broadcast wasted latency. To get a robust comparison, we run 10 experiments with random graph to get the 25th, median, 75th percentile of $i$-th node performance. The last epoch (99) of Perigee is shown as the dashed line in the figure. 
Fig.~\ref{fig:goldfish_perf} shows $\GF$ keeps improving performance until it converges at a region around epoch 64 and 99; $\GF$ has better performance than Perigee.

To robustly compare $\GF$ and Perigee, we run both systems under various network scenario, and each scenario is repeated with 10 different network initialization. 
Table~\ref{table:table:rand_g_perf} records the performance at their final epochs after summarizing over 10 runs. We use the identical evaluation method as above to show broadcast wasted latency. 
As the number of adapting nodes increase, both systems produce better performance, but $\GF$ is consistently better in all categories. We also observed that when there are many adaptation nodes, $\GF$ is only slightly better than the Perigee; the difference is most significant when around $30\%$ nodes in the network are adapting.

\subsection{Evaluation based on real world latency}
We use real world geolocation and latency to setup the network. The result is shown in the lower part of Table~\ref{table:table:rand_g_perf}. Compared to Perigee, $\GF$ on average uses $14.5\%$ less time. 




\section{Conclusion and Discussion}
P2P is a fundamental component in many decentralized applications. Our work addresses improvement on broadcast latency and exploration convergence by bringing idea from matrix completion and streaming algorithm to synthesize a general view about the network. When the number of publisher is equal or less than the number of exploitation connections, the $\GF$ can find and retain the global optimal connections. When the network is dynamic and complex, $\GF$ saves $14.5\%$ time than its baseline.

 \bibliography{references.bib}

\begin{thebibliography}{10}
\providecommand{\url}[1]{#1}
\csname url@samestyle\endcsname
\providecommand{\newblock}{\relax}
\providecommand{\bibinfo}[2]{#2}
\providecommand{\BIBentrySTDinterwordspacing}{\spaceskip=0pt\relax}
\providecommand{\BIBentryALTinterwordstretchfactor}{4}
\providecommand{\BIBentryALTinterwordspacing}{\spaceskip=\fontdimen2\font plus
\BIBentryALTinterwordstretchfactor\fontdimen3\font minus
  \fontdimen4\font\relax}
\providecommand{\BIBforeignlanguage}[2]{{%
\expandafter\ifx\csname l@#1\endcsname\relax
\typeout{** WARNING: IEEEtran.bst: No hyphenation pattern has been}%
\typeout{** loaded for the language `#1'. Using the pattern for}%
\typeout{** the default language instead.}%
\else
\language=\csname l@#1\endcsname
\fi
#2}}
\providecommand{\BIBdecl}{\relax}
\BIBdecl

\bibitem{mao2020perigee}
Y.~Mao, S.~Deb, S.~B. Venkatakrishnan, S.~Kannan, and K.~Srinivasan, ``Perigee:
  Efficient peer-to-peer network design for blockchains,'' in \emph{Proceedings
  of the 39th Symposium on Principles of Distributed Computing}, 2020, pp.
  428--437.

\bibitem{dembo2020everything}
A.~Dembo, S.~Kannan, E.~N. Tas, D.~Tse, P.~Viswanath, X.~Wang, and O.~Zeitouni,
  ``Everything is a race and nakamoto always wins,'' in \emph{Proceedings of
  the 2020 ACM SIGSAC Conference on Computer and Communications Security},
  2020, pp. 859--878.

\bibitem{flashbot}
\BIBentryALTinterwordspacing
flashbot. the-cost-of-resilience. [Online]. Available:
  \url{https://writings.flashbots.net/the-cost-of-resilience/}
\BIBentrySTDinterwordspacing

\bibitem{thedefiant}
\BIBentryALTinterwordspacing
thedefiant. flashbots-tornado-sanctions-mev. [Online]. Available:
  \url{https://thedefiant.io/flashbots-tornado-sanctions-mev}
\BIBentrySTDinterwordspacing

\bibitem{vyzovitis2020gossipsub}
D.~Vyzovitis, Y.~Napora, D.~McCormick, D.~Dias, and Y.~Psaras, ``Gossipsub:
  Attack-resilient message propagation in the filecoin and eth2.0 networks,''
  2020.

\bibitem{maymounkov2002kademlia}
P.~Maymounkov and D.~Mazieres, ``Kademlia: A peer-to-peer information system
  based on the xor metric,'' in \emph{International Workshop on Peer-to-Peer
  Systems}.\hskip 1em plus 0.5em minus 0.4em\relax Springer, 2002, pp. 53--65.

\bibitem{el2003efficient}
S.~El-Ansary, L.~O. Alima, P.~Brand, and S.~Haridi, ``Efficient broadcast in
  structured p2p networks,'' in \emph{International workshop on Peer-to-Peer
  systems}.\hskip 1em plus 0.5em minus 0.4em\relax Springer, 2003, pp.
  304--314.

\bibitem{dabek2004vivaldi}
F.~Dabek, R.~Cox, F.~Kaashoek, and R.~Morris, ``Vivaldi: A decentralized
  network coordinate system,'' \emph{ACM SIGCOMM Computer Communication
  Review}, vol.~34, no.~4, pp. 15--26, 2004.

\bibitem{augustine2022fully}
J.~Augustine, S.~Chatterjee, and G.~Pandurangan, ``A fully-distributed scalable
  peer-to-peer protocol for byzantine-resilient distributed hash tables,'' in
  \emph{Proceedings of the 34th ACM Symposium on Parallelism in Algorithms and
  Architectures}, 2022, pp. 87--98.

\bibitem{goldfish}
\BIBentryALTinterwordspacing
``Goldfish library.'' [Online]. Available:
  \url{https://github.com/bx3/goldfish}
\BIBentrySTDinterwordspacing

\bibitem{ripeanu2001peer}
M.~Ripeanu, ``Peer-to-peer architecture case study: Gnutella network,'' in
  \emph{Proceedings first international conference on peer-to-peer
  computing}.\hskip 1em plus 0.5em minus 0.4em\relax IEEE, 2001, pp. 99--100.

\bibitem{pytorch}
\BIBentryALTinterwordspacing
Pytorch. [Online]. Available: \url{https://pytorch.org/docs/stable/index.html}
\BIBentrySTDinterwordspacing

\bibitem{softmax}
\BIBentryALTinterwordspacing
wikipedia. softmax. [Online]. Available:
  \url{https://en.wikipedia.org/wiki/Softmax\_function}
\BIBentrySTDinterwordspacing

\bibitem{wonderproxy}
\BIBentryALTinterwordspacing
wondernetwork. wondernetwork. [Online]. Available:
  \url{https://wonderproxy.com/blog/a-day-in-the-life-of-the-internet/}
\BIBentrySTDinterwordspacing

\end{thebibliography}

\end{document}